\documentclass[aps, prd, nofootinbib, preprintnumbers, showpacs, showkeys,preprint]{revtex4}
\usepackage{graphicx}
\usepackage{amsmath}
\usepackage[usenames]{color}
\usepackage{allpurpose}

\begin{document}

\title{The four-momentum conservation and equal velocity assumption in neutrino oscillations}

\author{Junji Jia}
\email{jjia5@uwo.ca}
\affiliation{Department of Applied
Mathematics, University of Western Ontario, London, Ontario N6A 5B7,
Canada}


\date{\today}
\vspace{2cm}

\begin{abstract}

The neutrino masses and oscillation are closely related to how neutrino propagates. In this paper, we first derive a new form of four-momentum conservation that connects the four-momentum $(E_{\alpha},~p_{\alpha})$ of the neutrino flavor eigenstate and the four momenta $(E_i,~p_i)$ of its mass eigen-components. Then we use the assumption that the mass eigen-components travel at equal velocity to derive the energy-momentum square difference (EMSD) $E_{\alpha}^2-p_{\alpha}^2$ for the flavor eigenstate. It is shown that for the equal velocity assumption, $E_{\alpha}^2-p_{\alpha}^2$ will be a fixed constant in events with different $E_{\alpha}$ and/or $p_{\alpha}$. In contrast, for the equal energy or equal momentum assumption, $E_{\alpha}^2-p_{\alpha}^2$ will vary with $E_{\alpha}$ or $p_{\alpha}$ respectively. These EMSD can be checked by ongoing and future neutrino experiments. The phase difference between two neutrino mass eigen-components $|\nu_i\rangle$ and $|\nu_j\rangle$ is then derived for the equal velocity assumption. For relativistic neutrinos, it is shown that the phase difference depends linearly on the distance energy ratio $L/E$, the mass difference $m_i-m_j$ and an effective mass, in contrary to the linear dependance on $L/E$ and $m_i^2-m_j^2$ as the phase difference for equal energy or equal momentum assumption dose. The new phase difference implies different bounds on the neutrino masses or mass difference.

\end{abstract}



\maketitle

\section{Introduction}

The neutrino oscillations recently observed for the atmospheric neutrinos \cite{atm}, solar neutrinos \cite{solar} and reactor neutrinos \cite{reactor} have produced an explosion of both the experimental and theoretical studies of the neutrino. These observations confirmed that the neutrino flavor eigenstates are different from their mass eigenstates by a unitary transform
\be |\nu_{\alpha}\rangle = \sum_{i}U^*_{\alpha i}|\nu_i\rangle\label{transf}\ee
where $|\nu_{\alpha}\rangle$ denotes the flavor eigenstate of generation $\alpha$, $|\nu_{i}\rangle$ denotes the mass eigenstate with mass $m_i$ and $U$ is a unitary matrix. Information about the neutrino mass, mass bounds and mass differences were deduced from these experiments and they are used in particle physics \cite{Bhattacharyya:2003gi}, muon decay parameters \cite{Erwin:2006uc} and cosmology \cite{King:2002js}. In order to refine these observations and then put more stringent limit on the neutrino masses, more neutrino experiments are being conducted or proposed \cite{Tayloe:2003fd, Harris:2005yb, Ichikawa:2010zz}.

The key relation that is used to link the oscillation and masses is the phase difference $\Delta\Phi_{ij}$ between different mass eigenstates
\be \Delta\Phi_{ij}\equiv (E_it_i-p_ix_i) -(E_jt_j-p_jx_j) .\label{phased}\ee
$\Delta\Phi_{ij}$ can be further simplified to depend on the masses $m_k$. By fitting the experimentally observed transition probability $P_{\nu_\alpha\to\nu_\beta}$, which is a function of the phase difference $P_{\nu_\alpha\to\nu_\beta}=P_{\nu_\alpha\to\nu_\beta}\lb\Delta\Phi_{ij}(m_k)\rb$, one obtains information about $m_k$'s and their differences or bounds.

The phase difference \refer{phased} can be recast into the following form \cite{Lipkin:2005kg}
\be \Delta\phi_{ij}=\frac{\Delta m_{ij}^2}{p_i+p_j}\cdot L , \label{lipkinphase}\ee
where $\Delta m_{ij}^2\equiv m_i^2-m_j^2$ and \mcl denotes the distance from the neutrino source to the detector. It is often assumed that all mass eigen-components have {\it equal energy} $E_i=E$ \cite{Lipkin:1995cb,Grossman:1996eh,Stodolsky:1998} or {\it equal momentum} $p_i=p$ \cite{Bilenky:1978nj,Bilenky:2006zq}, then for ultra-relativistic neutrinos with $E_i\approx p_i\gg m_i$, this phase difference becomes the ``standard'' one that is used in most literatures \cite{Lipkin:1995cb,Grossman:1996eh,Stodolsky:1998,Bilenky:1978nj,Bilenky:2006zq,Amsler:2008zzb}
\be \Delta \phi_{ij}\approx \frac{L}{E}\cdot \frac{\Delta m_{ij}^2}{2}. \label{spd}\ee

The equal energy assumption (EEA) or equal momentum assumption (EMA) that is used to derive \refer{spd} however has no direct experimental justification. Theoretically, these two assumptions are not Lorentz invariant: when one boost from the frame in which they are valid to other frames, the energies or momenta of components with different masses will not remain the same \cite{Giunti:2001kj}. In order for \refer{spd} to be useful to us, the EEA and EMA will have to be valid in our laboratory frame for all neutrino sources/experiments, which is very unlikely, even though still possible. Due to this reason, these two assumptions are not specially favored over other possible assumptions \footnote{It is argued that there exist approaches that do not use the equal energy or momentum assumption \cite{Winter:1981kj, Giunti:2008cf, Cohen:2008qb}. However because their results for the phase difference is the same as the ``standard'' ones \refer{spd}, which is different from what we will derive using the equal velocity assumption (EVA), we believe it is still worthy to consider the EVA.}. Motivated by seeking a Lorentz invariant description, in this paper we examine some consequences of the {\it equal velocity assumption} (EVA); that is, all mass eigen-components of one flavor eigenstate propagate at equal velocity. The Lorentz invariance of this assumption is shown in the appendix \ref{appd}.

In section \ref{fmcsec}, we first derive a relation between the four-momenta of a flavor eigenstate and its mass eigen-components using the four-momentum conservation. Then using the EVA, we derive in section \ref{dispsec} the energy-momentum relation for the neutrino flavor eigenstate. It will be seen that the energy-momentum square difference (EMSD) $E_{\alpha}^2-p_{\alpha}^2$ derived from the EVA is a constant for any flavor neutrino $|\nu_\alpha\rangle$ with energy $E_{\alpha}$ and momentum $p_{\alpha}$, while those EMSD derived from EEA and EMA varies with the energy and momentum respectively. Because of this difference, these assumptions have an opportunity to be experimentally discriminated, even though the accuracy requirement might be challenging. In section \ref{pdsec}, the phase difference between two mass eigen-components is found to have a different dependance on the mass eigenvalues $m_i$, compared to the usual phase difference \refer{spd}. Bounds on the masses $m_i$ and the mass difference are also calculated. In these derivations, we explicitly work in the case that there are three or two generation of flavor eigenstates, but the results are extendable to more than three generations. In section \ref{discsec}, we discuss the pion decay and the possible kinematical and dynamical reasons for the EVA.

\section{Four-momentum conservation}\label{fmcsec}
Properties of neutrinos, especially their masses, are mostly measured in experiments using decays of particles involving neutrinos. Among these, the tritium beta decay, the pion decay and the tau decay are the kinematically easiest. In tritium decay experiment, one measures the energy spectrum of electrons near the end-point. The shape of end-point of this spectrum is affected by the mass of electron neutrino in a known way and therefore can be used to deduce the electron neutrino mass \cite{Kraus:2004zw,Lobashev:1999tp}. In pion decay experiment, muon and muon neutrino are produced. Measuring the momentum of muon in the rest frame of pion, the kinematics of such a two body decay allows one to solve the muon neutrino mass directly \cite{Assamagan:1995wb}. The kinematics of tau decay through channel $\tau \to 3 \pi/5 \pi+\nu_{\tau}$ can similarly put an upper bound on the mass of tau neutrinos \cite{Barate:1997zg}.

For concreteness, let us consider the following example. Suppose that a neutrino flavor eigenstate $|\nu_{\alpha}\rangle$ is detected in an event of the form
\be \sum_n |\mbox{particle}_n\rangle\to|\nu_{\alpha}\rangle.\label{event}\ee
This can include the tritium, pion and tauon decay, and other neutrino events if we put every other particle except the neutrino to the left of the arrow. Further let us suppose that the four-momentum $p^{\mu}_{\alpha}$ of this neutrino flavor eigenstate $|\nu_{\alpha}\rangle$ is experimentally deducible from measurement of four-momenta of other particles and using the four-momentum conservation law in this event, i.e.,
\be \sum_n p^{\mu}_n=p_{\alpha}^{\mu}\equiv(E_{\alpha},\vp_{\alpha})\label{fmconv}\ee
where $p^{\mu}_n$ is the four-momentum of $|\mbox{particle}_n\rangle$.
Supposing the three mass eigen-components $|\nu_i\rangle$ in the flavor eigenstate $|\nu_{\alpha}\rangle$ have four-momenta $p^{\mu}_i=(E_i,\vp_i)$ and velocities as $\vv_i$ for $i=1,~2,~3$ respectively, then these quantities should satisfy the on-shell condition
\be E_i^2=p_i^2+m_i^2, \ee where $p_i\equiv |\vp_i|$ and the relativistic four-momentum-velocity relation
\be E_i=\gamma_i m_i,\quad \vp_i=\gamma_i m_i\vv_i \label{vmr} \ee where we take $c=1$ and $\gamma_i=1/\sqrt{1-v_i^2}$ is the Lorentz factor of $|\nu_i\rangle$, $v_i\equiv |\vv_i|$ is the speed of $|\nu_i\rangle$.

We claim that these four-momenta satisfy the four-momentum conservation law of the following form
\be  p^{\mu}_{\alpha}=\sum_i |U_{\alpha i}|^2p^{\mu}_i\label{epconv}\ee or explicitly
\bea E_{\alpha}&=&\sum_{i} |U_{\alpha i}|^2 E_i,\label{econv} \\
\vp_{\alpha}&=&\sum_{i} |U_{\alpha i}|^2 \vp_i. \label{pconv}\eea
This form of four-momentum conservation is based on the following argument. Suppose we have a four-momentum operator $P^{\mu}$ that is applicable to the particles on the left hand side of event \refer{event} and on the mass eigen-component $|\nu_i\rangle$ to yield their four-momenta respectively
\bea P^{\mu}|\mbox{particle}_n\rangle &=&p^{\mu}_n|\mbox{particle}_n\rangle,\label{pparticlea}\\
P^{\mu}|\nu_i\rangle&=&p^{\mu}_i |\nu_i\rangle.\label{pparticleb}\eea
Evaluating this operator on both sides of event \refer{event} and using the orthonormal relations
\bea \langle\mbox{particle}_n|\mbox{particle}_m\rangle &=&\delta_{mn}~,\label{ortha}\\
\langle\nu_i|\nu_j\rangle&=&\delta_{ij}~,\label{orthb}\eea
we can find the four-momentum of the flavor eigenstate $|\nu_\alpha\rangle$
\be \begin{array}{rcll} p^{\mu}_\alpha&=&\displaystyle\sum_n p^\mu_n &\mbox{ (using definition \refer{fmconv})}\\
&=&\displaystyle\sum_n\langle\mbox{particle}_n|P^\mu|\mbox{particle}_n\rangle &\mbox{ (using \refer{pparticlea} and \refer{ortha})}\\
&=&\displaystyle\lb\sum_n\langle\mbox{partial}_n|\rb P^\mu\lb\sum_m|\mbox{particle}_m\rangle\rb &\mbox{ (using \refer{ortha})}\\
&=&\displaystyle\langle\nu_\alpha|P^\mu|\nu_\alpha\rangle &\mbox{ (using \refer{event})}\\
&=&\displaystyle\lb \sum_iU_{\alpha i}\langle \nu_i|\rb P^\mu\lb \sum_jU^*_{\alpha j}|\nu_j\rangle\rb &\mbox{ (using \refer{transf})}\\
&=&\displaystyle\sum_i|U_{\alpha i}|^2p^\mu_i &\mbox{ (using \refer{pparticleb} and \refer{orthb})}. \end{array}\ee
Therefore the above claim \refer{epconv}, which is just the four-momentum conservation law \refer{fmconv}, is justified.

This form of the four-momentum conservation is based on the ability to assign a separate four-momentum to each mass eigen-components of the flavor neutrino, and on the existence of a four-momentum operator $P^{\mu}$. They are fair assumptions that are used explicitly or implicitly in most literatures. Therefore we believe that the conservation law \refer{epconv} derived from these assumptions is correct and we will use it in the following sections. In the appendix, we also show that this form of four-momentum conservation is compatible with Lorentz boosts.

\section{Equal velocity assumption}
Now we make our main assumption of this paper: the velocities of three mass eigen-components are equal; that is, they have the same magnitude and direction
\be \vv_1=\vv_2=\vv_3=\vv.\label{equalv}\ee
This assumption has the advantage that it is Lorentz invariant, i.e., it holds in any reference frame. In the appendix, this invariance is explicitly shown. Moreover, in order for the interference phenomenon, such as neutrino oscillation, to happen, different components must remain coherent. Because all mass eigen-components are generated at the same space point, the EVA will guarantee that they stick together all the time and therefore remain coherent and the neutrino can oscillate.

In the following sections, we will show that this assumption has interesting and experimentally checkable consequences. But before we do that, it seems appropriate here to address an argument against the EVA that was raised in Refs. \cite{Okun:2000hm,Okun:2000gc} and discussed in \cite{Levy:2000qx}. The argument is based on the following expressions. Using relation \refer{vmr}, the assumption \refer{equalv} immediately implies that
\be \frac{E_i}{E_j}=\frac{\gamma m_i}{\gamma m_j}=\frac{m_i}{m_j},\qquad \frac{p_i}{p_j}=\frac{\gamma m_iv}{\gamma m_jv}=\frac{m_i}{m_j}. \label{evaeq}\ee Then the argument was that while $m_i/m_j$ may be very small or large, in real experiment such as the $\pi^+$ decay, $E_i/E_j$ is always approximately 1. However, it seems this argument is not valid here. In the neutrino detectors, the energy of neutrino is always {\it derived} from the measurement of other particles in the same electroweak processes (using equation \refer{fmconv} in events like \refer{event}). Thus what is measured in these experiments is always the energy of a neutrino flavor eigenstate, $E_\alpha$, but no the energy of its mass-eigencomponents, $E_i$. And what is approximately one is the energy ratio $E_{\alpha 1}/E_{\alpha 2}$ of two neutrino flavor eigenstates in two events, but not the energy ratio of two components. Therefore we believe the argument in Ref. \cite{Okun:2000hm,Okun:2000gc} will not invalidate the EVA.

\subsection{Energy-Momentum Square Difference}\label{dispsec}

The first of the consequences that follows from the EVA is a constant EMSD $E_{\alpha}^2-p_{\alpha}^2$ for any flavor neutrino $|\nu_\alpha\rangle$. Using relation \refer{vmr} and the EVA \refer{equalv}, the energy \refer{econv} and the momentum \refer{pconv} become the following
\bea && E_{\alpha}=\sum_i |U_{\alpha i}|^2 \gamma m_i , \label{energycon}\\
     && p_{\alpha}=\sum_i |U_{\alpha i}|^2 \gamma m_i v ,\label{momentumcon}\eea where $v=|\vv|$, $\gamma=1/\sqrt{1-v^2}$.
Subtracting square of \refer{energycon} and \refer{momentumcon}, we get an energy-momentum relation
\be E_{\alpha}^2-p_{\alpha}^2=\lb\sum_i |U_{\alpha i}|^2 m_i\rb^2.\label{disp}\ee

Because $U_{\alpha i}$ and $m_i$ are fixed physical constants, the equality \refer{disp} implies that $E_{\alpha}^2-p_{\alpha}^2$ is Lorentz invariant, i.e., it has the same value in any reference frame. The Lorentz invariance is also expected from the four-momentum conservation law \refer{fmconv}: since the momenta $p_n^{\mu}$ of particles on the left hand side of \refer{event} are Lorentz covariant, the momentum $p^{\mu}_{\alpha}$, which is just a sum of $p^\mu_n$, is also Lorentz covariant. However, equality \refer{disp} further implies $E_{\alpha}^2-p_{\alpha}^2$ is single-valued. Note that the Lorentz invariance of $E_{\alpha}^2-p_{\alpha}^2$ dose not guarantee it is also single-valued: in two events with neutrino four-momenta $(E_{\alpha1}, p_{\alpha1})$ and $(E_{\alpha2},p_{\alpha2})$, each of $E_{\alpha1}^2-p_{\alpha1}^2$ and $E_{\alpha2}^2-p_{\alpha2}^2$ will be Lorentz invariant but there is no a priori condition that forces them to be the same value, given that in our theory a neutrino flavor eigenstate is a superposition of three mass eigen-components.

It is this single-valueness of the EMSD \refer{disp} that allows us to experimentally discriminate the EVA from the EEA and EMA. Consider the EEA first. Under this assumption, all mass eigen-components have energy $E$. The energy conservation \refer{econv} then gives \be E=E_{\alpha}.\ee The momentum of component $|\nu_i\rangle$ then reads
\be p_i=\sqrt{E_{\alpha}^2-m_i^2}.\ee
Using the momentum conservation \refer{pconv} we get
\be p_{\alpha}=\sum_i|U_{\alpha i}|^2 \sqrt{E_{\alpha}^2-m_i^2}. \ee
Therefore for the EEA, we get the EMSD
\be E_{\alpha}^2-p_{\alpha}^2=E_{\alpha}^2-\lb \sum_i|U_{\alpha i}|^2 \sqrt{E_{\alpha}^2-m_i^2}\rb^2. \label{star1}\ee
In the general case, the square of the sum in the right hand side of this equation dose not factor our $E_{\alpha}^2$ and the entire right hand side will vary with the variation of $E_{\alpha}$. Therefore in an experiment if two events with four-momenta $(E_{\alpha1}, p_{\alpha1})$ and $(E_{\alpha2},p_{\alpha2})$ were measured to have different energy $E_{\alpha1}\neq E_{\alpha2}$, the EEA will yield $E_{\alpha1}^2-p_{\alpha1}^2\neq E_{\alpha2}^2-p_{\alpha2}^2$, while the EVA have  $E_{\alpha1}^2-p_{\alpha1}^2 = E_{\alpha2}^2-p_{\alpha2}^2$. Similarly for the EMA, if each component has momentum $p$, then from \refer{pconv},
\be p=p_{\alpha},\ee
and therefore using \refer{econv}
\be E_{\alpha}=\sum_i|U_{\alpha i}|^2\sqrt{p_{\alpha}^2+m_i^2},\ee
and finally we get
\be E_{\alpha}^2-p_{\alpha}^2=\lb \sum_i|U_{\alpha i}|^2\sqrt{p_{\alpha}^2+m_i^2} \rb^2-p_{\alpha}^2. \label{star2}\ee
In the general case the right hand side has a non-trivial dependence on $p_{\alpha}$, which means $E_{\alpha}^2-p_{\alpha}^2$ will also be different for two events with different momentum $p_{\alpha}$.

Under current experimental conditions, only in the pion decay $\pi^+\to \mu^++\nu_{\mu}$ can one directly deduce $(E_{\nu_{\mu}},p_{\nu_{\mu}})$ from the measurement of muon energy and momentum. However the $E_{\nu_{\mu}}^2-p_{\nu_{\mu}}^2$ was found negative and therefore not physical \cite{Assamagan:1995wb}. The accuracy of this experiment will have to be improved and other experiments are needed in order to check the EMSD \refer{disp}, \refer{star1} and \refer{star2} and discriminate which assumption is correct. If the EVA is correct and future experiments can measure the four-momentum $(E_{\alpha},p_{\alpha})$ for all three neutrino flavors $|\nu_{\alpha}\rangle$, then the relation \refer{disp} can be directly inverted to find out $m_i$, assuming that $|U_{\alpha i}|^2$'s are independently measurable. These masses can be cross-checked with the masses obtained from neutrino oscillation experiments, which directly measure some function of masses $m_i$. For the ``standard'' phase difference \refer{spd}, this function is $\Delta m_{ij}^2$.

\section{Phase Difference and Mass Bounds}\label{pdsec}
Here we show that using the EVA, the phase difference defined in \refer{phased} will take a slightly different form from \refer{spd}. Dividing \refer{momentumcon} by \refer{energycon}, we get \be v=\frac{p_{\alpha}}{E_{\alpha}}.\label{vsol}\ee
Now using \refer{vsol}, $p_k=\gamma m_kv$, $E_k=\gamma m_k$ $(k=i,j)$ and $\gamma=1/\sqrt{1-v^2}$, the phase difference \refer{phased} becomes
\be  \Delta\phi_{ij}=\frac{m_i-m_j}{\displaystyle \frac{p_{\alpha}}{E_{\alpha}}}\sqrt{1-\frac{p_{\alpha}^2}{E_{\alpha}^2}}\cdot L=\frac{L}{p_{\alpha}}\cdot \Delta m_{ij} \sum_k|U_{\alpha k}|^2m_k,\label{ph2}\ee
where $\Delta m_{ij}\equiv m_i-m_j$ and in the last step the EMSD \refer{disp} is used. It is convenient to define an effective mass $m^{\mbox{\footnotesize eff}}_{\alpha}$ for the flavor eigenstate $|\nu_{\alpha}\rangle$
\be m^{\mbox{\footnotesize eff}}_{\alpha}\equiv \sqrt{E_{\alpha}^2-p_{\alpha}^2}=\sum_i |U_{\alpha i}|^2 m_i\label{effm}.\ee The phase difference \refer{ph2} then becomes \be \Delta\phi_{ij}=\frac{L}{p_\alpha}\cdot m^{\mbox{\footnotesize eff}}_{\alpha}\Delta m_{ij}\label{ph2rew}.\ee

The result \refer{ph2} or \refer{ph2rew} is exact so far. It is valid to both the relativistic and non-relativistic neutrinos.  If one further assume that the neutrino is ultra-relativistic, i.e., $E_{\alpha}/p_{\alpha}\approx 1$, then the phase difference \refer{ph2} becomes
\be \Delta\phi_{ij}\approx\frac{ L}{E_\alpha} \cdot \Delta m_{ij} \sum_k|U_{\alpha k}|^2m_k. \label{phrel}\ee This form of the phase difference has the same dependence as the ``standard'' phase difference \refer{spd} on the source-detector distance $L$ and the energy of neutrino $E_{\alpha}$. However they have an important difference in the remaining factor, i.e., the part besides $L/E$ in the phase differences. The phase difference \refer{spd} depends linearly on $\displaystyle \frac{\Delta m_{ij}^2}{2}=(m_i-m_j)\lb\frac{m_i}{2}+\frac{m_j}{2}\rb$ while the phase difference \refer{ph2} depends linearly on $\displaystyle\Delta m_{ij}\sum_k|U_{\alpha k}|^2m_k=(m_i-m_j)\lb |U_{\alpha 1}|^2m_1+\cdots+|U_{\alpha q}|^2m_q\rb$ where $q$ is the number of flavors.
This factor is one main parameter the neutrino oscillation experiments measure. From this factor, which henceforth will be denoted as $f^2/2$ ($f>0$) to simplify our later calculation,  and using different phases \refer{spd} and \refer{ph2}, bounds of different values on the neutrino masses can be extracted.

This can be most clearly seen if we work in an situation that there are only two neutrino flavors and if we express $m_i$ and $m_j$ in terms of $\Delta m_{ij}$ and $f$. For the ``standard'' phase difference \refer{spd}, we can solve the definitions
\be   \frac{1}{2}(m_i+m_j)(m_i-m_j)=\frac{f^2}{2}, \quad m_i-m_j=\Delta m_{ij} \ee to obtain
\bea m_i&=&\frac{1}{2}\lb\frac{f^2}{\Delta m_{ij}}+\Delta m_{ij}\rb,\\
m_j&=& \frac{1}{2}\lb\frac{f^2}{\Delta m_{ij}}-\Delta m_{ij}\rb. \eea
Without losing any generality, we can assume that $m_i\geq m_j$ and therefore $\Delta m_{ij}\geq 0$. Then the positivity of $m_j$ and $\Delta m_{ij}$ implies
\bea \Delta m_{ij}&\leq &f,\\
\mbox{ and }m_i&\geq &f.\label{clb}\eea
For the new phase difference \refer{ph2}, we have
\be (m_i+m_j)(|U_{\alpha i}|^2 m_i+|U_{\alpha j}|^2 m_j)=\frac{f^2}{2}, \quad m_i-m_j=\Delta m_{ij}, \ee
from which we solve
\bea m_i&=&\frac{1}{2} \frac{f^2}{\Delta m_{ij}}+|U_{\alpha j}|^2\Delta m_{ij} ,\\
m_j&=& \frac{1}{2} \frac{f^2}{\Delta m_{ij}}-|U_{\alpha i}|^2\Delta m_{ij} . \eea
Then the same positivity of $m_j$ and $\Delta m_{ij}$ only implies \bea
\Delta m_{ij}&<&\frac{1}{|U_{\alpha i}|}\frac{f}{\sqrt{2}},\\
m_i&>&\lcb\ba{ll}|U_{\alpha j}|\sqrt{2}f&\mbox{ when }|U_{\alpha j}|>|U_{\alpha i}|\\
\displaystyle \frac{1}{|U_{\alpha i}|}\frac{f}{\sqrt{2}}&\mbox{ when }|U_{\alpha j}|\leq|U_{\alpha i}|\ea
\right., \eea where we implicitly used $|U_{\alpha i}|^2+|U_{\alpha j}|^2=1$. When $|U_{\alpha i}|$ is small, the upper bound on $\Delta m_{ij}$ can be very large. The lower bound on $m_i$ can be smaller or larger by a factor of $\sqrt{2}$ than the corresponding bound \refer{clb} predicted by the ``standard'' phase difference. Because of the importance of these bounds in particle physics \cite{Bhattacharyya:2003gi}, muon decay parameters \cite{Erwin:2006uc} and cosmology \cite{King:2002js}, it becomes important to check which phase difference (\refer{spd} or  \refer{ph2}) is correct and whether neutrino mass-eigencomponents propagate at equal-velocity.

\section{Discussion}\label{discsec}
We have seen that one important consequence of the EVA is the single-valueness of $E_{\alpha}^2-p_{\alpha}^2$, or equivalently $m_{\alpha}^{\mbox{\footnotesize eff}}$. Here we consider the single-valueness of $E_{\alpha}^2-p_{\alpha}^2$ in the pion decay $\pi^+\to \mu^++\nu_{\mu}$ and argue that it favors the EVA. Let us consider pion decay in the rest frame of the pion. Using the four-momentum conservation \refer{fmconv} in this case, we have \bea E_{\nu_{\mu}}&=& m_{\pi}-\sqrt{p_{\mu}^2+m_{\mu}^2}\\
p_{\nu_{\mu}}&=&p_{\mu}\eea where $p_{\mu}$ is the momentum of the muon and $m_{\pi}$ and $m_{\mu}$ are the masses of the pion and the muon respectively. Because experiment only measures one value of $p_{\mu}$ \cite{Assamagan:1995wb}, $(E_{\nu_{\mu}},p_{\nu_{\mu}})$ and therefore also $E_{\nu_{\mu}}^2-p_{\nu_{\mu}}^2$ have only one value in this frame. As argued in section \ref{dispsec}, this fact favors the EVA over the EEA and EMA. If one takes a more aggressive point of view by insisting that the EEA and EMA should not only {\it allow} but also {\it require} the appearance of multiple values of $E_{\nu_{\mu}}^2-p_{\nu_{\mu}}^2$ if one repeats the experiment many times -- since there is no other kinematical reasons that stops the multiple values of the EMSD from appearing for the EEA and EMA -- then the pion decay experiment can even exclude the EEA and EMA. Note that in this experiment the $E_{\nu_{\mu}}^2-p_{\nu_{\mu}}^2$ is just the effective mass square $m_{\nu_{\mu}}^2$ of the muon neutrino flavor eigenstate $|\nu_{\mu}\rangle$ \cite{Assamagan:1995wb}.

If we extend the single-valueness of $E_{\nu_\mu}^2-p_{\nu_\mu}^2$ of $m^{\mbox{\footnotesize eff}}_{\nu_\mu}$ in the pion decay and further require that $m^{\mbox{\footnotesize eff}}_{\alpha}$ of flavor eigenstate $|\nu_{\alpha}\rangle$ is indeed independent of the velocities of its mass eigen-components,
\be \frac{\dd}{\dd v_j}\lb E_{\alpha}^2-p_{\alpha}^2\rb =
\frac{\dd }{\dd v_j}\lsb \lb \sum_i|U_{\alpha i}|^2\gamma_im_i\rb^2-\lb \sum_i|U_{\alpha i}|^2\gamma_im_iv_i\rb^2\rsb =0,\quad (j=1,2,3). \label{vind}\ee
Then it is readily to find that in the general case that $m_i$'s are not degenerate or zero, the only solution to \refer{vind} is $v_1=v_2=v_3$. Therefore the EVA can be thought as a kinematical consequence of the independence requirement of $E_{\alpha}^2-p_{\alpha}^2$ on $v_j$'s. However to avoid any misleading, we need to point out that this velocity independence requirement is only a sufficient but not necessary condition for the single-valueness of the $E_{\alpha}^2-p_{\alpha}^2$ or $m_{\alpha}^{\mbox{\footnotesize eff}}$.

If we accept the EVA, then the most important question becomes: what is the dynamical reason for the equal velocity? In this analysis, we implicitly assumed that there are three mass components which, besides that they are linearly superposed to form a flavor eigenstate, are dynamically independent. However in reality they might interact through weak processes and this interaction might provide all/part of the reason for the EVA. Note that one can try to ask the same question for quarks too: Do quark mass eigen-components of a flavor eigenstate propagate at the same velocity? There in the strong interactions, if quarks are expressed in mass eigen-components then the CKM matrix has to be incorporated. However because there is no free quark propagating due to confinement, the question how dose a free quark flavor eigenstate travel in vacuum can be avoided.

Our treatment of neutrino in this work depends primarily on the ability to assign a four-momentum and velocity to each mass eigen-component. Therefore our treatment is classical: we do not consider any quantum fluctuation of the particle \cite{Lipkin:2005kg}. If a more rigorous wave packet treatment were used and we still want to impose the same EVA, then we would have to assign the central values of the four-momentum of the wave packet $|\nu_i\rangle$ as $p^{\mu}_i$ in \refer{vmr} and require it to satisfy the relation \refer{evaeq}, which is equivalent to the EVA in this case. Indeed, the wave packet treatment in Ref. \cite{Takeuchi:2000fz} (the eqs. (3.5) and (3.6)) would have led to our phase difference \refer{ph2} if they are granted the four-momentum conservation law \refer{epconv} and the EVA.


Finally let us point out that the EVA in neutrino oscillation was first suggested in \cite{Takeuchi:1998kx} and then considered in \cite{DeLeo:1999aa, Takeuchi:2000fz, Okun:2000hm, Okun:2000gc, Field:2002gg}. However none of the EMSD \refer{disp}, \refer{star1} or \refer{star2} or the phase difference \refer{ph2rew} or \refer{phrel} was derived. Indeed, these references got a phase difference that is similar to the ``standard'' one \refer{spd}. Therefore the results in this paper are all new, in addition to that they are experimentally checkable.

\begin{acknowledgments}
This work was supported by the Natural Sciences and Engineering Research Council of Canada.
\end{acknowledgments}

\appendix*
\section{Equal velocity assumption, four-momentum conservation and Lorentz boosts} \label{appd}
In this appendix we show that the EVA and the four-momentum conservation \refer{fmconv} are compatible with Lorentz boosts.

Suppose the four-momentum of a flavor eigenstate is measured in one frame as $(E_{\alpha},p_\alpha)$ (for simplicity let us assume that all momenta are in a fixed direction). And suppose in the same reference frame the four-momenta and velocities of the mass eigen-components are $(E_i,p_i)$ and $v_i$ respectively. Let us suppose the Lorentz boost has a relative velocity $u$ in the same direction of the motion of the neutrino with respect to the initial frame and let us denote $\beta^\prime=u/c$ and $\gamma^\prime =1/\sqrt{1-\beta^{\prime 2}}$. Under this boost, the four-momenta $(E_i,p_i)$ change to $(E_i^\prime, p_i^\prime)$ in the following way
\be \lb \ba{c}E_i^\prime \\ p_i^\prime\ea\rb =\gamma^\prime \lb \ba{cc}1&-\beta^\prime\\ -\beta^\prime & 1\ea\rb \lb \ba{c}E_i\\ p_i\ea\rb\,,\quad i=1\,,2\,,3\,.\ee
Explicitly, they are
\bea E_i^\prime &=& \gamma^\prime (E_i-\beta^\prime p_i)\\
p_i^\prime &=& \gamma^\prime (-\beta^\prime E_i+p_i).\eea

If the energies of different mass eigen-components were equal: $E_i=E_j$, then because $m_i\neq m_j$ and hence $p_i\neq p_j$ (or $E_i\neq E_j$), we will have
\be E_i^\prime = \gamma^\prime (E_i-\beta^\prime p_i)\neq  \gamma^\prime (E_j-\beta^\prime p_j)=E_j^\prime. \ee
Similarly, if the momenta of different mass eigen-components were equal: $p_i=p_j$, then because $m_i\neq m_j$ and hence $E_i\neq E_j$, we will have
\be p_i^\prime = \gamma^\prime (-\beta^\prime E_i+p_i)\neq \gamma^\prime (-\beta^\prime E_j+p_j)=p_j ^\prime.\ee
It is seen then the equal energy or momentum assumption is not compatible with Lorentz boosts. While for the equal velocities: $v_i=v_j$, under the same boost they transform to
\be v_i^\prime =\frac{v_i-u}{\displaystyle 1-\frac{v_iu}{c^2}}. \ee Apparently, the same velocity assumption holds under this boost.

Now we show that the four-momentum conservation law also holds under the boost. Because the four-momentum $(E_{\alpha},p_{\alpha})$ is deduced from event \refer{event} or formula \refer{fmconv}, then under the Lorentz boost the requirement of four-momentum conservation is equivalent to the Lorentz invariance of $E_{\alpha}^2-p^2_\alpha$. The later could be shown easily. Using the four-momentum relation \refer{econv} and \refer{pconv}, we have
\bea && E^\prime_\alpha=\sum_i|U_{\alpha i}|^2 E_i^\prime =\sum_i|U_{\alpha i}|^2 \gamma^\prime (E_i-\beta^\prime p_i)=\gamma^\prime (E_\alpha-\beta^\prime p_\alpha) \\
&& p^\prime_\alpha=\sum_i|U_{\alpha i}|^2 p_i^\prime =\sum_i|U_{\alpha i}|^2 \gamma^\prime (-\beta^\prime E_i+p_i)=\gamma^\prime (-\beta^\prime E_\alpha+p_\alpha). \eea
It is then readily to check the Lorentz invariance of $E_\alpha^2-p_\alpha^2$ \bea E_{\alpha}^{\prime 2}-p_{\alpha}^{\prime 2}&=&\gamma^{\prime 2}\lsb E_{\alpha}^2(1-\beta^{\prime 2})-(1-\beta^{\prime 2})p_{\alpha}^2\rsb \\
&=& E_\alpha^2-p^2_\alpha. \eea

The above proves that the four-momentum conservation law of the form \refer{fmconv} is compatible with Lorentz boosts.


\begin{thebibliography}{99}

\bibitem{atm}
  Y.~Ashie {\it et al.}  [Super-Kamiokande Collaboration],
  Phys.\ Rev.\ Lett.\  {\bf 93}, 101801 (2004)
  [arXiv:hep-ex/0404034].
  Phys.\ Rev.\  D {\bf 71}, 112005 (2005)
  [arXiv:hep-ex/0501064].

\bibitem{solar}
  S.~Fukuda {\it et al.}  [Super-Kamiokande Collaboration],
  Phys.\ Rev.\ Lett.\  {\bf 86}, 5656 (2001)
  [arXiv:hep-ex/0103033].
  Phys.\ Lett.\  B {\bf 539}, 179 (2002)
  [arXiv:hep-ex/0205075].
  B.~Aharmim {\it et al.}  [SNO Collaboration],
  Phys.\ Rev.\  C {\bf 72}, 055502 (2005)
  [arXiv:nucl-ex/0502021].
  J.~Hosaka {\it et al.}  [Super-Kamkiokande Collaboration],
  Phys.\ Rev.\  D {\bf 73}, 112001 (2006)
  [arXiv:hep-ex/0508053].

\bibitem{reactor}
  M.~H.~Ahn {\it et al.}  [K2K Collaboration],
  Phys.\ Rev.\  D {\bf 74}, 072003 (2006)
  [arXiv:hep-ex/0606032].
  D.~G.~Michael {\it et al.}  [MINOS Collaboration],
  Phys.\ Rev.\ Lett.\  {\bf 97}, 191801 (2006)
  [arXiv:hep-ex/0607088].

\bibitem{Bhattacharyya:2003gi}
  G.~Bhattacharyya, H.~Pas, L.~g.~Song and T.~J.~Weiler,
  Phys.\ Lett.\  B {\bf 564}, 175 (2003)
  [arXiv:hep-ph/0302191].

\bibitem{Erwin:2006uc}
  R.~J.~Erwin, J.~Kile, M.~J.~Ramsey-Musolf and P.~Wang,
  Phys.\ Rev.\  D {\bf 75}, 033005 (2007)
  [arXiv:hep-ph/0602240].

\bibitem{King:2002js}
  S.~F.~King,
  arXiv:hep-ph/0210089.
  A.~D.~Dolgov,
  Surveys High Energ.\ Phys.\  {\bf 17}, 91 (2002)
  [arXiv:hep-ph/0208222].


\bibitem{Tayloe:2003fd}
  R.~Tayloe  [MiniBooNE Collaboration],
  Nucl.\ Phys.\ Proc.\ Suppl.\  {\bf 118}, 157 (2003).

\bibitem{Harris:2005yb}
  D.~A.~Harris  [MINOS and NOvA Collaborations],
  Nucl.\ Phys.\ Proc.\ Suppl.\  {\bf 149}, 150 (2005).

\bibitem{Ichikawa:2010zz}
  A.~K.~Ichikawa  [T2K Collaboration],
  J.\ Phys.\ Conf.\ Ser.\  {\bf 203}, 012104 (2010).
  R.~Terri  [T2K Collaboration],
  Nucl.\ Phys.\ Proc.\ Suppl.\  {\bf 189}, 277 (2009).

\bibitem{Lipkin:2005kg}
  H.~J.~Lipkin,
  Phys.\ Lett.\  B {\bf 642}, 366 (2006)
  [arXiv:hep-ph/0505141], see equation (1.4).

\bibitem{Lipkin:1995cb}
  H.~J.~Lipkin,
  Phys.\ Lett.\  B {\bf 348}, 604 (1995)
  [arXiv:hep-ph/9501269];
  arXiv:hep-ph/9901399;
  Phys.\ Lett.\  B {\bf 579}, 355 (2004)
  [arXiv:hep-ph/0304187].

\bibitem{Grossman:1996eh}
  Y.~Grossman and H.~J.~Lipkin,
  Phys.\ Rev.\  D {\bf 55}, 2760 (1997)
  [arXiv:hep-ph/9607201].

\bibitem{Stodolsky:1998}
  L.~Stodolsky,
  Phys.\ Rev.\  D {\bf 58}, 036006 (1998)
  [arXiv:hep-ph/9802387].


\bibitem{Bilenky:1978nj}
  S.~M.~Bilenky and B.~Pontecorvo,
  Phys.\ Rept.\  {\bf 41}, 225 (1978).

\bibitem{Bilenky:2006zq}
  S.~M.~Bilenky and M.~D.~Mateev,
  Phys.\ Part.\ Nucl.\  {\bf 38}, 117 (2007)
  [arXiv:hep-ph/0604044].

\bibitem{Amsler:2008zzb}
  C.~Amsler {\it et al.}  [Particle Data Group],
  Phys.\ Lett.\  B {\bf 667}, 1 (2008).

\bibitem{Giunti:2001kj}
  C.~Giunti,
  Mod.\ Phys.\ Lett.\  A {\bf 16}, 2363 (2001)
  [arXiv:hep-ph/0104148];
  arXiv:hep-ph/0105319.

\bibitem{Winter:1981kj}
  R.~G.~Winter,
  Lett.\ Nuovo Cim.\  {\bf 30}, 101 (1981).

\bibitem{Giunti:2008cf}
  C.~Giunti,
  AIP Conf.\ Proc.\  {\bf 1026}, 3 (2008)
  [arXiv:0801.0653 [hep-ph]].

\bibitem{Cohen:2008qb}
  A.~G.~Cohen, S.~L.~Glashow and Z.~Ligeti,
  Phys.\ Lett.\  B {\bf 678}, 191 (2009)
  [arXiv:0810.4602 [hep-ph]].

\bibitem{Kraus:2004zw}
  C.~Kraus {\it et al.},
  Eur.\ Phys.\ J.\  C {\bf 40}, 447 (2005)
  [arXiv:hep-ex/0412056].

\bibitem{Lobashev:1999tp}
  V.~M.~Lobashev {\it et al.},
  Phys.\ Lett.\  B {\bf 460}, 227 (1999).

\bibitem{Assamagan:1995wb}
  K.~Assamagan {\it et al.},
  Phys.\ Rev.\  D {\bf 53}, 6065 (1996).

\bibitem{Barate:1997zg}
  R.~Barate {\it et al.}  [ALEPH Collaboration],
  Eur.\ Phys.\ J.\  C {\bf 2}, 395 (1998).

\bibitem{Okun:2000hm}
  L.~B.~Okun,
  Surveys High Energ.\ Phys.\  {\bf 15}, 75 (2000).

\bibitem{Okun:2000gc}
  L.~B.~Okun and I.~S.~Tsukerman,
  Mod.\ Phys.\ Lett.\  A {\bf 15}, 1481 (2000)
  [arXiv:hep-ph/0007262].

\bibitem{Levy:2000qx}
  J.~M.~Levy,
  arXiv:hep-ph/0012285.


\bibitem{Takeuchi:1998kx}
  Y.~Takeuchi, Y.~Tazaki, S.~Y.~Tsai and T.~Yamazaki,
  Mod. Phys. Lett. A {\bf 14}, 2329, (1999)
  [arXiv:hep-ph/9809558].

\bibitem{DeLeo:1999aa}
  S.~De Leo, G.~Ducati and P.~Rotelli,
  Mod.\ Phys.\ Lett.\  A {\bf 15}, 2057 (2000)
  [arXiv:hep-ph/9906460].

\bibitem{Takeuchi:2000fz}
  Y.~Takeuchi, Y.~Tazaki, S.~Y.~Tsai and T.~Yamazaki,
  Prog.\ Theor.\ Phys.\  {\bf 105}, 471 (2001)
  [arXiv:hep-ph/0006334].

\bibitem{Field:2002gg}
  J.~H.~Field,
  Eur.\ Phys.\ J.\  C {\bf 30}, 305 (2003)
  [arXiv:hep-ph/0211199].

\end{thebibliography}
\end{document}